
\input harvmac
\def\dual{\kern-.25em\,^*\kern-.20em}
\def\ext{{\rm d}}
\def\bh{black hole}
\def\df{degrees of freedom}
\def\Hu{H^{\mu\nu\rho}}\def\Hl{H_{\mu\nu\rho}}
\def\hu{h^{\mu\nu\rho}}\def\hl{h_{\mu\nu\rho}}
\def\ep{\epsilon^{\mu\nu\rho\lambda}}
\def\Bl{B_{\mu\nu}}
\def\mpla{{\sl Mod. Phys. Lett.}}
\def\st{space-time}\def\td{\widetilde D}
\def\Sp{$\Sigma$-projection}
\def\IR{{\rm I}\kern-.20em{\rm R}}
\def\vq{\vec q}
\def\Q{Q_{\rm rms}}
\noblackbox
\Title{LA-UR-92-471}{An Alternative Scenario for Non-Abelian Quantum Hair}
\centerline{Amitabha Lahiri\footnote{$^{\dag}$}{(lahiri@pion.lanl.gov)}}
\bigskip\centerline{Theoretical  Division  T-8}
\centerline{Los Alamos National Laboratory}
\centerline{Los Alamos, NM 87545}
\vskip0.3in
\centerline{\bf Abstract}

Topologically charged \bh s in a theory with a 2-form coupled
to a non-abelian gauge field are investigated. It is found that the
classification of the ground states is similar to that in the theory of
non-abelian discrete quantum hair.

\Date{02/92}
\newsec{Introduction}

Until recently it was thought that the powerful classical `no-hair' theorems
\ref\wald{R. Wald, {\it General Relativity and Gravitation}, Addison-Wesley,
1982.} always forced an isolated \bh\ to radiate away almost all of
its internal information, retaining only a handful of states characterized
by its mass $M$, angular momentum $J$ and charges associated with massless
gauge fields. However, since a \bh\ may be expected to form by absorbing
matter originally in a pure quantum state and to decay by predominantly
thermal radiation, it would seem that information is irretrievably lost in
the process of formation and subsequent decay of a \bh. Unless, that is,
the \bh\ can support a large number of `hair' or internal \df\ that
correlate the radiated states with the information that went into its
formation. This contradiction has been a source of consternation to all who
believe that quantum coherence is preserved in all processes. In addition,
a stable \bh\ may be shown to have a very large entropy, proportional to
the area of its event horizon. This also suggests that a \bh\ should be able to
access a very large number of internal \df.  In recent years, new hopes of
resolving this conflict have been aroused by the emergence of the idea of
`quantum hair' \ref\CPW{A lucid account of the present state of affairs may
be found in {\sl Quantum Hair on Black Holes} by S. Coleman, J. Preskill
and F. Wilczek, IASSNS-HEP-91/64.}. That a \bh\ can carry topological
charge and therefore global degrees of freedom not fully described by
massless gauge fields was first postulated in
\ref\bghhs{M. J.  Bowick {\sl et al.}, {\sl Phys. Rev. Lett.} {\bf
61} (1988) 2823.}. Even though most of the analysis was classical, the only
apparent way of detecting the charge was via an Aharonov-Bohm experiment
that required a closed fundamental string whose world sheet fully enclosed
the horizon of the \bh. The topological charge, corresponding to the
integral of a 2-form field $B$ over a closed 2-surface around the \bh,
introduced a phase factor in the wave-function of the string and its effect
was therefore quantum in nature. Another type of hair was discovered soon
afterwards \ref\KW{L. M. Krauss and F. Wilczek, {\sl Phys. Rev.  Lett.} {\bf
62} (1989) 1221.}, where the quantum nature of the hair, arising from the
spontaneous breaking of a local continuous symmetry to a local discrete
symmetry, was more evident. The `hair' in this case is the charge
associated with the local discrete group -- the part of the continuous
symmetry charge that is not screened by spontaneous symmetry breaking. An
understanding of a possible relation between the two types of hair was
non-existent until it was shown \ref\abl{T. J. Allen, M. J. Bowick and A.
Lahiri, {\sl Phys. Lett.} {\bf 237B} (1989) 47.} that if the $B$ field is
coupled to a $U(1)$ gauge field $A$ via an $mB\wedge F$ term (where $F=\ext
A$), the resulting theory is a `dual' description of the Goldstone
mechanism. Since the `screening' of charge through symmetry breaking
is a crucial ingredient in the theory of discrete quantum hair, it seems
plausible \CPW\ that the two theories are actually
different descriptions of the same physical phenomenon. However, a rigorous
proof of a one-to-one correspondence between the states of the two theories
has not yet been found.

In view of this, one is tempted to look for a non-abelian generalization of
the $B$ field where the corresponding topological charge may relate to the
discrete charge one obtains by breaking an $SU(N)$ group to a local
discrete symmetry group. Such a generalization will be considered in what
follows, and we will find that the topological charges on \bh s in such a
theory are indeed very interesting.

\newsec{The Non-Abelian 2-Form}

We begin by briefly summarizing some features of the abelian $B$ field. The
$B$ field is a 2-form potential, {\it i.e.}, an antisymmetric tensor field
of rank 2 with `field strength' $H$ defined by $\Hl =
\del_{[\mu}B_{\nu\rho]}$. (We assume that the space-time connection is
torsion-free,
so $\grad\mu$ can be replaced by $\del_\mu$ in antisymmetric derivatives,
as in the definition of $H$.)  One insists that an action involving $B$ be
invariant under a Kalb-Ramond symmetry transformation, $\Bl\to\Bl +
\del_{[\mu}\Lambda_{\nu]}$. When coupled to a gauge field $A$ that has an
associated abelian gauge symmetry under $A_\mu\to A_\mu +
\del_\mu\lambda$, the action that is formally invariant under
both symmetry transformations is given by\foot{Here and later on, the
presence of the Einstein-Hilbert term and the Einstein equations are
understood.}
\eqn\abac{S = \int\big(\sqrt{-g}(-{1\over4}F_{\mu\nu}F^{\mu\nu}
-{1\over12}\Hl\Hu)
+ {m\over2}\ep F_{\mu\nu}B_{\rho\lambda}\big).}
The equations of motion following from this action are\foot{The
usual convention is to write the last term in \abac\ as $mB\wedge F \equiv
{m\over 4}\ep F_{\mu\nu}B_{\rho\lambda}$. Here we try to keep the equations
neat and absorb the factor of 2 in $m$.}
\eqn\abeom{\eqalign{\grad\nu F^{\nu\mu} + m\ep H_{\nu\rho\lambda} &= 0,\cr
\grad\rho\Hu + m\ep F_{\rho\lambda} &= 0.\cr}}
These are the well known London equations of superconductivity, and it can
be shown \nref\jmrw{J. Minahan and R. Warner, {\sl Stuckelberg Revisited},\
University of Florida preprint UFIFT-HEP-89-15.}
\nref\ablii{T. J. Allen, M. J. Bowick and A. Lahiri, \mpla\ {\bf
A6} (1991) 559.} that for appropriate gauge choices these equations would
lead to massive equations for either $A$ or $B$ \abl-\ablii. In the context
of \bh s one can show that these equations force both $F$ and $H$ to vanish
outside the horizon of a static, spherically symmetric \bh. It then follows
that the vacuum Einstein equations hold outside the horizon, and \bh\
uniqueness theorems can be used to prove that the space-time metric is
Schwarzschild, while the \bh\ may carry a topological charge,
\eqn\topocharge{B = q\varpi \equiv
{q\over{4\pi}}\sin\theta\ext\theta\wedge\ext\phi.}
In components $B_{\theta\phi} = {q\over{4\pi r^2}}$, and one should note
that $B$ can be thought of as a long range field even when it gains a mass
via the $mB\wedge F$ term.
It would
be interesting to see if similar results hold in the presence of a
non-abelian gauge field to which a `non-abelian' $B$ couples, and if we are
led to a non-abelian topological charge.

Our starting point is a naive non-abelianization of the action \abac,
\eqn\nabac{S = \int\Tr\big(\sqrt{-g}(-{1\over4}F_{\mu\nu}F^{\mu\nu}
-{1\over12}\Hl\Hu) + {m\over 2}\ep F_{\mu\nu}B_{\rho\lambda}\big),}
where we assume that $B$ belongs to the adjoint representation of SU(N) with
$B_{\mu\nu}\to UB_{\mu\nu}U^{-1}$ and $A_\mu\to UA_\mu U^{-1} - \del_\mu
UU^{-1}$ under a gauge transformation with $U\in$ SU(N).  The relevant field
strengths are defined as $\Hl = D_{[\mu}B_{\nu\rho]} \equiv
\del_{[\mu}B_{\nu\rho]} + [A_{[\mu}, B_{\nu\rho]}]$, and $F_{\mu\nu} =
[D_\mu, D_\nu]$. (It should be
noted that the covariant transformation law of $B$ does not come from
`first principles', but is assumed in order to leave this particular action
invariant. There are known theories containing an antisymmetric tensor
that does not transform covariantly, or even has a well-defined local
gauge transformation law \ref\FT{E. S. Fradkin and A. A. Tseytlin, {\sl
Annals of Physics} {\bf 162} (1985) 31.}.) Obviously, one cannot naively
non-abelianize
the Kalb-Ramond symmetry with these definitions, because under
$B_{\mu\nu}\to B_{\mu\nu} + D_{[\mu}\Lambda_{\nu]}$ the action is not
invariant. The Kalb-Ramond symmetry is responsible in the abelian case for
the equivalence of $B$ with a Goldstone boson, and losing this symmetry
evidently has serious consequences.
(For an action that retains a non-abelian Kalb-Ramond symmetry by
introducing auxiliary fields, see \ref\Raj{S. G. Rajeev, {\sl
Duality and Gauge Invariance}, MIT preprint CTP-1335.}.)
However, not all is lost. We will
examine the question of symmetries later; first we investigate the results
that follow from the action \nabac.

The equations of motion following from \nabac\ are
\eqn\aeom{D_\nu F^{\nu\mu} - [B_{\nu\rho}, \Hu] + m\ep H_{\nu\rho\lambda}
= 0,}
and
\eqn\beom{D_\rho\Hu + m\ep F_{\rho\lambda} = 0.}
Operating with $D_\mu$ on \aeom\ gives
\eqn\consi{- [B_{\nu\rho}, D_\mu\Hu] + m\ep[F_{\mu\nu}, B_{\rho\lambda}]= 0,}
which contains no new information (it is the commutator of \beom\ with $B$),
while operating on \beom\ first with $D_\nu$ and then with $D_\mu$ gives us
\eqn\consii{\eqalign{[F_{\nu\rho}, \Hu] &= 0, \cr
[F_{\nu\rho}, D_\mu\Hu] &= 0.\cr}}
It follows from these equations that the non-abelian $A$-$B$ system is highly
constrained. In
fact, an analysis of the constraints in the system shows that there are a
total of two dynamical \df\ left in the theory. A simple physical argument
can be used to count the \df\ as follows. One notes that when $m = 0$, these
equations demand that $H = 0$ for a generic $F$, and this sector of solutions
is the same as standard gauge theory. If the
space-time has trivial second cohomology (as is thought to be the case in
all known experiments of particle physics) the topological modes
of $B$ can be gauged away\foot{See section 3.}. As $m$ is turned on, the term
$mB\wedge F$ induces a mixing
between the modes of $A$ and $B$. But since this is a topological term in
the action (it can be defined without the help of a background metric), it
has vanishing contribution to the energy momentum tensor. Therefore it does
not contribute to the energy-momentum carried by propagating modes, and
also cannot generate an extra mode. This is very different from the abelian
case, where setting $m = 0$ completely decouples the two fields, and the
resulting theory has three \df\ to start with. In the non-abelian case the
equations \consii\ show that there is no non-zero $H$ that is independent
of $F$, even when $m = 0$.

At this point however, we are interested in the physics generated by
\nabac\ in the context of \bh\ space-times. Our analysis will follow
closely that of \abl, which was done for abelian gauge fields. Consider a
static \st, {\it i.e.}, one with a hypersurface-orthogonal timelike Killing
vector field $\xi^\mu$, $\biglie_\xi g_{\mu\nu} = 0$, $\xi_\mu\xi^\mu = -
\lambda^2$.  Let $\Sigma$ denote a hypersurface to which $\xi$ is
orthogonal, and $\Pi^\mu_{\mu'} = \delta^\mu_{\mu'} +
\lambda^{-2}\xi^\mu\xi_{\mu'}$ the projection operator that projects into
$\Sigma$. Also let $\Omega$ be a $p$-form on the \st\ manifold $M$ and
$\omega$ its projection on $\Sigma$ with $\biglie_\xi\Omega = 0$. Denoting
the induced connection on $\Sigma$ by $\widetilde\nabla$, it can be shown that
\eqn\lemma{\widetilde\nabla_\alpha(\lambda\omega^{\alpha\mu\cdots\nu})
= \lambda\Pi^\mu_{\mu'}\cdots\Pi^\nu_{\nu'}\grad\alpha\Omega^{\alpha\mu'\cdots
\nu'}.}
Physically this may be understood as saying that the 3-divergence of a form
is equal to its 4-divergence when all fields (including the metric) are
time-independent. Let $\td_\mu$ denote the \Sp\ of the gauge-covariant
derivative operator $D_\mu$. It follows from \lemma\ that
\eqn\nlemma{\td_\alpha(\lambda\omega^{\alpha\mu\cdots\nu}) =
\lambda\Pi^\mu_{\mu'}\cdots\Pi^\nu_{\nu'}D_\alpha\Omega^{\alpha\mu'\cdots
\nu'} + \lambda^{-1}\xi^\alpha\xi_{\alpha'}[A_{\alpha},
\Omega^{\alpha'\mu\cdots\nu}].\footnote{\ddag}{Note that in a gauge
$\xi^\alpha A_\alpha = 0$ the last term vanishes. Such a gauge choice
is always possible because
$\xi_\alpha\Pi_{(A)}^\alpha\approx 0$ is a constraint of the theory. However,
it
is not necessary to choose a gauge at this point.}}

Let us denote the \Sp s of $H$, $F$, $\dual H$ and $\dual F$ by $h$, $f$, $d$
and $e$, respectively. Then by \nlemma\ and by the equations of motion, we have
\eqn\nproj{\eqalign{\td_\rho(\lambda\hu) &= -\lambda me^{\mu\nu}
+ \lambda^{-1}\xi^{\rho'}\xi_\rho[A_{\rho'}, \Hu] ,\cr
\td_\nu(\lambda f^{\nu\mu}) &= \lambda md^\mu + \lambda[B_{\nu\rho},
H^{\mu'\nu\rho}]\Pi^\mu_{\mu'} +
\lambda^{-1}\xi^\nu\xi_{\nu'}[A_\nu, F^{\nu'\mu}].\cr}}
We multiply the first of these equations by $e_{\mu\nu}$, take the trace
and integrate over the region $V$ between the horizon and the sphere at
infinity (we have assumed a spherically symmetric \st, $M\simeq
S^2\times\IR^2$). This gives us
\eqn\trint{\int_V\Tr\Bigg[e_{\mu\nu}\bigg(\td_\rho(\lambda\hu) + \lambda
me^{\mu\nu} - \lambda^{-1}\xi^{\rho'}\xi_\rho[A_{\rho'}, \Hu]\bigg)\Bigg] = 0.}
Integrating by parts and using assumptions of regularity of the horizon
(field strengths are finite when $\lambda$ = 0) and asymptotic flatness
(field strengths $\to$ 0 as $r \to \infty$) and using the \Sp s of the
equations of motion \aeom\ and \beom\ we obtain
\eqn\nproji{\int_V\Tr\bigg(\lambda m\hl\hu + \lambda me_{\mu\nu}e^{\mu\nu}
+ \lambda[ B_{\mu\nu}, (\dual H)_\rho]\hu
-\lambda^{-1}\xi^{\rho'}\xi_\rho[A_{\rho'}, \Hu]e_{\mu\nu}\bigg) = 0.}
 Similarly, by multiplying the
second equation of \nproj\ by $d_\mu$, taking the trace and going through a
similar calculation one finds that
\eqn\nprojii{\int_V\Tr\bigg(\lambda mf_{\mu\nu}f^{\mu\nu} + \lambda md_\mu
d^\mu + \lambda[ B_{\nu\rho}, \Hu]d_\mu +
\lambda^{-1}\xi^\nu\xi_{\nu'}d_\mu[A_\nu, F^{\nu'\mu}]\bigg) = 0.}
A little algebra shows that when the two equations \nproji\ and \nprojii\ are
added, the last terms cancel each other\foot{Here we have chosen
a gauge $\xi^\mu B_{\mu\nu} = 0$, which we are always allowed to do as
$\xi_\mu\Pi_B^{\mu\nu} \approx 0$ is a constraint of the theory.
Note that the solutions in sec. 3 are
consistent with this gauge choice.}, and one is left with
\eqn\vanish{\int_V\lambda\Tr(\hl\hu + e_{\mu\nu}e^{\mu\nu}
+ f_{\mu\nu}f^{\mu\nu} + d_\mu d^\mu) = 0.}
Since the metric is positive definite on the hypersurface $\Sigma$, this
equation implies that $d$,
$e$, $f$, $h$ all vanish. It follows that in fact $F$ and $H$ must vanish
outside the horizon of a static, spherically symmetric, asymptotically flat
\st. We will see that this result has very interesting topological
consequences for \bh\ vacua, which we define as a \st\ with a \bh\ with
vanishing field strengths outside the horizon.

\newsec{Solutions}

One possible solution to the equations $F = 0$, $H = 0$ is $A = 0$, $B =
\vq\varpi$, where $\varpi$ is again the harmonic form on the sphere, $\varpi
= \sin\theta\ext\theta\wedge\ext\phi$, while $\vq$ is now any vector in the Lie
algebra of SU(N). This can be said to have topological charge $\vq =
\int_{S^2}B$. This can be thought of as embedding an abelian group in SU(N),
corresponding to choosing a direction. However, the situation is not as
trivial as it seems, because this is not the only solution.
The vanishing of $F$
implies that at best $A$ is pure gauge, $A_\mu = - \del_\mu UU^{-1}$. Since
$H$ is covariant under SU(N) gauge transformations, it follows that $A_\mu
= - \del_\mu UU^{-1}$, $B = \vq(x)\varpi$ with $\vq(x) = U\vq U^{-1}$ is
also a solution to $F = 0$, $H = 0$. This solution evidently has a
different topological `charge', but can be thought of as an SU(N) gauge
transform of our first solution. However, something does remain invariant
under the gauge transformation -- it is the `magnitude' of the charge,
$|\vq(x)|^2 \equiv \Tr(\vq(x)\vq(x))$. As $\vq$ is an (N$^2$ --
1)-dimensional vector when $B$ belongs to the adjoint representation of
SU(N), the group of transformations that leave $|\vq(x)|^2$ invariant is
 O(N$^2$ -- 1).
Suppose we classify the
\bh\ vacua by the gauge-invariant `mean-square-charge' $\Q^2 =
\int_{S^2}|\vq(x)|^2$. Then for N $>$ 2  the
symmetry group of the \bh\ vacua is O(N$^2$ -- 1), which contains SU(N) as a
subgroup. It follows then, that different embeddings of SU(N) in O(N$^2$
-- 1) will provide gauge-inequivalent (in the sense of SU(N), but connected
by O(N$^2$ -- 1) transformations) \bh\ vacua, which may be classified by
the coset space O(N$^2$ -- 1)/SU(N).
For N = 2 this is not the correct description of \bh\ vacua. In this situation
the gauge group SU(2) is locally isomorphic to the symmetry group O(3) of the
\bh\ vacua, so there is no `manifold of vacua'. However, one may say that the
\bh\ vacua are now characterized by a winding number corresponding to
embeddings of SO(3)$\times$Z$_2$ in SU(2). The vacua with different winding
numbers are equivalent under SU(2) gauge transformations.

At this point we make a small digression and note that $A_\mu = 0$,
$B_{\mu\nu} = \del_{[\mu}\Lambda_{\nu]}$, where $\Lambda$ is a gauge-covariant
object ({\it i.e.,} $\Lambda \to U\Lambda U^{-1}$ under a gauge
transformation), is a trivial solution of the $F =
0 = H$ system, but the gauge transformed solution $A_\mu = - \del_\mu
UU^{-1}$, $B_{\mu\nu}(U) = U\del_{[\mu}\Lambda_{\nu]}U^{-1}$ is non-trivial
in the sense that $\int_{S^2}B(U)$ does not vanish identically. However, if
we write $\ext\Lambda = \vec
f(\theta,\phi)\sin\theta\ext\theta\wedge\ext\phi$ on the sphere for some
appropriate set of functions $f^a(\theta,\phi)$ ($a = 1,\cdots,N^2 - 1$)
such that
\eqn\trivial{\int_{S^2}f^a(\theta,\phi)\sin\theta\ext\theta\wedge\ext\phi =
0,}
it can be seen that the charge $\vq$ should be modified to $\vq + \vec
f(\theta,\phi)$, and then the modified mean-square-charge
$\widetilde\Q^2 = \int_{S^2}|\vq + \vec f(\theta,\phi)|^2$ is
invariant under SU(N). The group of invariance of $\widetilde\Q^2$ is again
O(N$^2$ -- 1), and our previous analysis of the classification of \bh\
vacua remains unaffected.

Coming back to the \bh\ vacuum states, one can compare the results obtained
here with similar results obtained for the theory of discrete quantum hair.
In the latter, one obtains discrete magnetic hair via the spontaneous
breakdown of SO(N$^2$ -- 1) to SU(N)/Z$_N$. One can then show \CPW\ that
the \bh\ vacua in the theory are connected by SO(N$^2$ -- 1)
transformations, but inequivalent under SU(N)/Z$_N$ transformations. The
coset space SO(N$^2$ -- 1)/(SU(N)/Z$_N$) then classifies the \bh\ vacuum
states. (We note that for N=2 this space gives the winding number of
SO(3) embeddings in SO(3), which is in essential agreement with the
findings above.)

\newsec{Conclusions}

It would seem that we have done no more than `almost' reproducing discrete
quantum hair via a rather unconventional route. We have reached a
classification scheme for \bh\ vacua that matches a similar classification
for a special class of discrete hair -- the discrete magnetic hair\foot{
In general, one hopes that SU(N) gauge theories confine the `electric' charge
and screen the `magnetic' charge, so this special class may be the only
one relevant to physics.}.
 So what have we gained by this analysis?

One important thing to note is that there is no scalar field involved in
this theory, $B$ is not dual to a scalar, unlike in the abelian case.  As a
result, one need not have a non-vanishing Higgs field in \st\ in order to
have non-abelian topological charge on a \bh. However, it is easy to see
that if the unbroken local symmetry group is SU(N)/Z$_N$, our analysis
reaches the same conclusions about the manifold of \bh\ vacua as one does
in the case of discrete magnetic hair. This seems to indicate that the $B$
field and discrete quantum hair may be describing very similar, but not
identical physical situations.

Another clue to the similarity or difference between the two theories may be
obtained by considering the experimental method of observing either
phenomenon. Discrete hair may be observed by `lassoing' a \bh\ by a
cosmic string loop that carries non-abelian electric (or magnetic) flux
along its core. A similar `lassoing', but only with a loop carrying
electric flux, would also detect the non-abelian $B$
field. In both processes, the loop picks up a phase proportional to the
product of the discrete or topological charge and the flux inside the
cosmic string core. This leads to an Aharonov-Bohm type effect.
In reality, both discrete magnetic hair and the $B$ charge couple to
a closed tube of electric flux. Such objects are unstable for SU(3) in the
presence of quarks. Since SU(3) is the group closest to experimental
physics for which one can obtain a non-trivial topological (or discrete)
charge, one should look for alternative
particle physics manifestations of topological
charge. The question of particle physics phenomena caused
by a $B$ field brings us to a question we have avoided answering so far --
Is there any symmetry (other than SU(N)) associated with the action \nabac?

The full symmetry of the action \nabac\ is at best poorly understood,
as there are a lot of technical difficulties associated with the quantization
of a non-abelian 2-form (for a discussion and a comprehensive list
of references, see \ref\bbt{D. Birmingham {\sl et al.}, {\sl Phys. Rep.}
{\bf 209} (1991) 129.}). However, the following may provide a pointer towards
relating the \nabac\ action to particle physics. As has been mentioned before,
a non-abelianization of the Kalb-Ramond gauge transformation, $\Bl\to\Bl +
D_{[\mu}\Lambda_{\nu]}$, does not leave $H$ invariant. However, there is
a `shift', or a field redefinition associated with $B$ that leaves $H$
invariant, $\Bl \to \Bl - \alpha F_{\mu\nu}$, where $\alpha$ is a dimensionful
parameter of mass dimension $-1$. In general, such a shift (similar to
the Goldstone mechanism for gauge fields, where the gauge field is `shifted'
by the derivative of a fundamental scalar already in the theory) leaves the
path-integral measure, and therefore amplitudes, invariant. Without going into
the subtleties of quantization, let us assume that the amplitudes remain
invariant in this theory as well. Let us now introduce an $F\dual F$ term
in the action,
\eqn\cpac{S = \int\Tr\big(\sqrt{-g} (-{1\over4}F_{\mu\nu}F^{\mu\nu}
-{1\over12}\Hl\Hu) + {m\over 2}\ep F_{\mu\nu}B_{\rho\lambda} + \Theta\ep
F_{\mu\nu}F_{\rho\lambda}\big).}
Evidently, by using the `shift' symmetry with $\alpha = 2\Theta/m$, one
can set the $\Theta$ angle to zero and be left with \nabac. Loosely
speaking, one may then expect the CP-violating effects of the gauge theory
to reside entirely in topological modes of $B$ (for \st s with trivial second
cohomology and in a state with finite action, one may write $B = U\ext\Lambda
U^{-1}$ at large $x$, as we found in the previous section) and not appear in
local
dynamics as they do via a $\Theta F\dual F$ term. The parameter $\alpha$
may be thought of as a constant or `invisible' axion\foot{Through
an unfortunate nomenclature, the $B$ field has often been confused with
an axion.
However, the only connection between the two (at least when $B$ is non-abelian)
seems to come from the above
speculation.}. We also note that the invariance of the action under this
shift symmetry may be used as a justification for the absence of other terms
involving $B$, if we maintain that bare couplings (other than $\Theta$)
should not be modified under a local symmetry of the action.
  More cannot be said without a deeper understanding of the
action \nabac, the shift symmetry, and its relevance to particle physics.
Work on this is in progress, and results will be reported elsewhere.

\centerline{\bf Acknowledgement}

It is a pleasure to thank T. J. Allen and M. J. Bowick for numerous
discussions leading to this work and for critically reading the manuscript.
Also many thanks to D. Daniel, N. Dorey, M. Mattis and especially E. Mottola
for many stimulating discussions.

\listrefs
\bye